\documentclass[12pt]{article}

\usepackage{answers}
\usepackage{setspace}
\usepackage{graphicx}
\usepackage{enumitem}
\usepackage{multicol}
\usepackage{mathrsfs}
\usepackage{appendix}
\usepackage[margin=2cm]{geometry} 
\usepackage{amsmath,amsthm,amssymb}
\usepackage{floatrow}
\usepackage{mathtools}  
\usepackage{commath}    
\usepackage{listings}
\usepackage{subfig}
\DeclarePairedDelimiter{\ceil}{\lceil}{\rceil}
\DeclarePairedDelimiter{\floor}{\lfloor}{\rfloor}

\usepackage{authblk}

\begin{document}
\setcounter{Maxaffil}{0}
\renewcommand\Affilfont{\itshape\small}
\title{Optimal assignment of buses to bus stops in a loop by reinforcement learning}
\author[1, 2]{Luca Vismara}
\author[2, 3, *]{Lock Yue Chew}
\author[2, 3]{Vee-Liem Saw}
\affil[1]{Interdisciplinary Graduate Programme, 61 Nanyang Drive, Nanyang Technological University, Singapore 637335}
\affil[2]{Division of Physics and Applied Physics, School of Physical and Mathematical Sciences, 21 Nanyang Link, Nanyang Technological University, Singapore 637371}
\affil[3]{Data Science and Artificial Intelligence Research Centre, Block N4 \#02a-32, Nanyang Avenue, Nanyang Technological University, Singapore 639798}
\affil[*]{Corresponding author: Lock Yue Chew, lockyue@ntu.edu.sg}

\maketitle

\begin{abstract}


Bus systems involve complex bus-bus and bus-passengers interactions. In this paper, we study the problem of assigning buses to bus stops to minimise the average waiting time of passengers. 
We formulate an analytical theory for two specific cases of interactions: the normal situation where all buses board passengers from every bus stop, versus the novel ``express buses" where disjoint subsets of non-interacting buses serve disjoint subsets of bus stops. Our formulation allows for the exact calculation of the average waiting time for general bus loops in the two cases examined. Compared with regular buses, we present scenarios where ``express buses" show an improvement in terms of average waiting time. From the theory we can obtain useful insights: 1) there is a minimum number of buses needed to serve a bus loop, 2) splitting a crowded bus stop into two less crowded ones always increases the average waiting time for regular buses, 3) changing the destination of passengers and location of bus stops do not influence the average waiting time. 
In the second part, we introduce a reinforcement-learning platform that can overcome the limitations of our analytical method to search for better allocations of buses to bus stops that minimise the average waiting time. Compared with the previous cases, any possible interaction between buses is allowed, unlocking novel emergent strategies. We apply this tool to a simple toy model and three empirically-motivated bus loops, based on data collected from the Nanyang Technological University shuttle bus system. In the simplified model, we observe an unexpected strategy emerging that could not be analysed with our mathematical formulation and displays chaotic behaviour. The possible configurations in the three empirically-motivated scenarios are approximately $10^{11}$, $10^{11}$ and $10^{20}$, so a brute-force approach is impossible. Our algorithm can reduce the average waiting time by $12\%$ to $32\%$ compared with regular buses and $12\%$ to $29\%$ compared with express buses. This tool can have practical applications because it works independently of the specific characteristics of a bus loop.

\end{abstract}

\begin{section}{Introduction}\label{subsec:busproblem}
%
%
%
%
%
Transportation planning in modern cities is an increasingly complex socio-economic issue because it involves many constraints such as infrastructural operating cost, passengers commuting time, and uncertain traffic conditions. The problem of a bus loop and in particular the problem of optimising the resources available (i.e. buses) to provide the best service for passengers has been studied extensively. 
This is important due to its implications in the everyday commute of people worldwide and its challenging complexity that may help in understanding similar problems involving the interaction of many agents. Service delays are detrimental to passengers \cite{Abkowitz1984} and incur an extra monetary cost for the company operating the line  \cite{Goeddel1996}. 

A common way to control buses is to employ dynamic control via holding strategies. These can be used to address the optimisation of waiting and travelling times \cite{Osuna1972}, to improve schedule reliability \cite{Abkowitz1984} \cite{Xuan2011} \cite{Chen2015} and to eliminate a common symptom of delays: bus bunching \cite{Daganzo2009} \cite{saw2019intelligent} \cite{Chew2019} \cite{Abkowitz1984} \cite{rossetti1998} \cite{Hickman2001} \cite{fu2002} \cite{cats2011} \cite{bartholdi2012} \cite{Moreira-Matias2016} \cite{quek2020analysis} \cite{saw2019}. An idea from Saw and Chew \cite{saw2019no} is to implement a no-boarding technique in place of the usual holding. In a scenario where alighting and boarding are sequential, a bus can disallow boarding to speed up. It has been shown from analytical calculations and numerical simulations how this can be effectively used to prevent bus bunching. The same authors also analyse the impact of such a policy in a game-theoretical model to verify its robustness and to suggest practical strategies to implement it  \cite{saw2020}. A more recent work by the authors  \cite{saw2019intelligent} uses reinforcement learning to optimise the headway between buses. Holding and no-boarding dynamical strategies emerge from simple stay/leave actions in an empirically-motivated scenario.

Some of the approaches in the literature obtain analytical results in idealised conditions. In an early work from Osuna and Newell  \cite{Osuna1972}, they study analytically a simplified model of a bus line implementing a dispatch or hold strategy to minimise the waiting time of passengers showing that, as expected, staggered buses deliver better performance. On the other hand, if the control is too tight, the system slows down and the effectiveness is reduced. More refined models have subsequently been developed \cite{Newell1974} \cite{Barnett1974} but one of the limitations of this approach lies in the threshold-based policy where a fixed threshold for the implementation of the holding control is obtained.
Another procedure to improve bus efficiency is to directly mitigate one of the main causes of delay and disruption, according to experience: bus bunching. Daganzo \cite{Daganzo2009} proposes a dynamical scheme to tackle this problem by using real-time headway information. This increases the efficiency in terms of holding time compared with fixed-threshold methods and it only considers the action of holding based on headway information.
Xuan, Argote and Daganzo  \cite{Xuan2011} introduce a family of dynamic holding strategies, obtained based on information on the leading bus, aimed at improving schedule adherence while keeping higher speed compared with the previous models based solely on headway information.

Chen, Chen and Chen \cite{Chen2015} propose a model that incorporates reinforcement learning. Their goal is to build a system that can carry out dynamic holding control in a noisy environment. The buses are modelled as agents and they minimise headway deviations. Using multi-agent reinforcement learning improves real-time operations compared with previous works which are focused on centralised control. The proposed model has a hierarchical approach since on top of the bus agents, other agents are needed to coordinate, manage and interface the agents with the environment.
A different approach is taken by Alesiani and Gkiotsalitis \cite{Alesiani2018}: they used Double Deep Q-learning \cite{Mnih2015} \cite{Hasselt:2016:DRL:3016100.3016191} to minimise deviations from a target headway, travel time and holding time. They combine these quantities in a customised cost function. The action used is holding the bus at a bus stop dynamically.

The need for dynamical adjustments is a common motif in all those works. The control depends on various parameters such as the positions of the buses or their deviations from the schedule, ultimately relying on the driver to take the right action at the right time. Until fully autonomous vehicles become mainstream, human error and potential distractions from instruction communicated in a dynamical situation can pose a safety risk. Another important commonality of these methods is that specific physical infrastructures are needed for holding and specific arrangements with passengers are needed when no-boarding actions are taken, which may limit the applicability of this class of techniques.

In contrast, we present the problem of how to distribute a number of buses to serve a number of bus stops. No dynamical intervention is considered, instead, the bus loop is optimised by restricting the buses at which passengers from a given bus stop can board. In other words, each bus boards passengers from a predetermined subset of the bus stops whilst alighting is always permitted. This makes it easy for commuters: as long as a bus is willing to pick them up, then they can alight where they wish to, which distinguishes this method from complete stop-skipping where alighting is also not allowed. In that case, the commuters must be aware to not board certain buses even if boarding is allowed from their origin bus stop. Depending on the configuration of the bus loop considered, a significant reduction in average waiting time can be expected. The easier implementation compared with dynamically controlled buses can open the possibility of a range of applications.

In section \ref{sec:cases} we formulate a theory of express buses and we compare them with regular uncontrolled buses in terms of average waiting time. The analytical calculations allow for general observations and recommendations on how to structure a bus loop, as summarised in section \ref{sec:insight}.
The calculations are possible because regular and express buses have simple interactions: with regular buses, all the buses interact with one other by serving the same bus stops and this causes bunching \cite{saw2019} \cite{Chew2019}.
Express buses instead operate on disjoint subsets of bus stops, eliminating the interaction between groups of buses. The interaction happens because the time spent serving a bus stop depends on the number of passengers waiting, which depends on when it was previously served by another bus. With regular and express buses this can be calculated.
To study the general case of arbitrary interaction between buses we introduce (section \ref{sec:MARL}) a minimal framework based on reinforcement learning; this generality opens up the possibility of real-world applications. We apply it to a toy model and three empirically-motivated bus loops and compare the solutions found with regular and express buses.
In the final part, in section \ref{sec:chaos}, a simple configuration discovered by the reinforcement learning algorithm, which shows chaotic behaviour, is briefly presented, based on Ref. \cite{saw2020chaos}.
The main findings are reported in section \ref{sec:conclusion} while the more technical details of the algorithm are in the appendix.



\end{section}

\section{Waiting time in different configurations of buses}\label{sec:cases}
Let us consider a loop with $M$ arbitrarily positioned bus stops: $\{S_i\}_{i=1}^M$ and $N$ buses: $\{B_i\}_{i=1}^N$. Passengers arrive at bus stop $S_i$ with constant rates $s_i \ge 0$ (passengers per unit time) and want to alight at a bus stop $S_j$ with probability $\zeta_{ij} \ge 0$, $\zeta_{ii} = 0$ and $\sum_{j=1}^M \zeta_{ij} = 1$, $\forall i$.
For simplicity, the number of passengers that can board or alight a bus per unit time are considered equal and defined as $l$. It is convenient to characterise a bus stop $S_i$ with the ratio between its arrival rate and the loading/unloading rate, to obtain the dimensionless parameters $k_i = s_i/l$. Buses are assumed to have a constant natural period $T$ over the loop, defined as time spent on the road excluding dwell time at bus stops. This means that all the buses move at the same speed on the road between bus stops. Acceleration and deceleration of buses are not considered and no constraint on the maximum capacity of the buses is imposed. Passengers alight first and then board. Buses are allowed to overtake.
In this loop, we want to compare the average waiting time for passengers in two cases:
\begin{enumerate}
\item Regular buses: uncontrolled buses serve all the bus stops;
\item Express buses: disjoint subsets of uncontrolled buses serve disjoint subsets of bus stops.
\end {enumerate}
In this context, ``uncontrolled'' means that the buses do not employ dynamic control.
The first case is what normally happens in many bus loops. It is well known that in such situation bunching occurs, which increases waiting time \cite{saw2019} \cite{Andres2017} \cite{Chen2016} \cite{Yu2016} \cite{Liang2016}. In the second case which we introduce in this paper, bunching between disjoint subsets of buses is avoided because the stopping time of a bus is independent of the position of the buses in different subsets. As the next section shows, no extra intervention from the driver or extra infrastructure is required once it is established that a certain bus will only serve a certain subset of the bus stops. Notice that if more than one bus serves the same subset of bus stops, partial bunching occurs, i.e. these buses which serve the same subset of bus stops would bunch into a single platoon.

The general idea to compute the average waiting time is to work under the assumption of constant arrival of passengers at a bus stop $i$ at a rate of $s_i$ passengers per unit time. Then, the average waiting time is half the longest possible waiting time \cite{saw2019no}. This is obtained by considering that all the waiting times are linearly decaying from the longest, up to $0$ for the passenger who arrives at the bus stop just before the bus leaves, hence boards without any delay. 

The presumption of constant arrival of passengers at a bus stop is convenient for our calculation but often the number of passenger waiting at a bus stop is modelled with a Poisson process \cite{Hickman2001} \cite{fu2002} \cite{cats2011} \cite{Moreira-Matias2016} \cite{quek2020analysis} \cite{saw2019no}. Through numerical simulations over the scenarios discussed in this paper, we observe that the formulas introduced in the next section for the average waiting time, Eqs. (\ref{eq:case1}) and (\ref{eq:case2}), are also valid if passengers arrive at each bus stop $i$ according to a Poisson process with $\lambda_i = s_i$. For this paper, however, we consider the arrival of passengers constant at the rate of $s_i$ passengers per unit time.

Ref. \cite{saw2020chaos} presents an analytical calculation for the average waiting time of regular and express buses whereby a bus stop is exclusively either an origin (passengers arrive at rate $s_i > 0$) or a destination (passengers want to alight there). The paradigm is different, but the average waiting time computed is the same as the result presented in this paper.

\subsection{Uncontrolled buses serving all the bus stops} \label{subsec:1}
In this case, it is well known that total bunching will occur, given enough time \cite{saw2019}. The calculation presented refers to the steady-state when all buses settle down into a fully bunched configuration.
In such configuration, the effective boarding/alighting rate of the buses is proportional to the number of buses so alighting and boarding are faster at a given bus stop. It is implicitly assumed that buses can board passengers in parallel when buses are at the same bus stop.
By considering as $\bar T = T + \sum_{i=1}^M\tau_i$ the total time taken for a bus to complete its loop, including the stopping time at bus stops (dwell time) for alighting and boarding of passengers $\{ \tau_i = \tau_i^{\text{alight}} + \tau_i^{\text{board}} \}_{i=1}^M$, the average waiting time for passengers boarding at bus stop $i$ is 
\begin{equation}\label{eq:wt}
\text{W}_i = \frac{\bar T - \tau_i^{\text{board}}}{2}. 
\end{equation}
Following Eq. (\ref{eq:wt})  what is left to compute is the dwell time spent for passengers to alight and board. In this configuration, for bus stop $i$ served by $N$ buses:
\begin{equation}\label{eq:tau}
\tau_i = \tau_i^{\text{alight}} + \tau_i^{\text{board}} = \frac{\sum_{j = 1}^M \zeta_{ji} k_j \bar T}{N} + \frac{k_i}{N} \bar T.
\end{equation}
The first term represents the time taken for the passengers to alight and depends on the number of passengers directed to the bus stop $i$ from other bus stops $j$. The second term is the time that it takes for passengers at bus stop $i$ to board the buses, which is proportional to the number of passengers to board $s_i \bar T$ (which also accounts for the passengers arriving during the alighting/boarding process) and inversely proportional to the effective boarding rate of $N$ bunched buses: $l N$.
This can be used to obtain $\bar T$ and compute the average waiting time for passengers arriving at bus stop $i$ from Eq. (\ref{eq:wt}):
\begin{equation}\label{eq:that}
\bar T = T +  \sum_{i=1}^M \left( \frac{\sum_{j = 1}^M \zeta_{ji} k_j \bar T}{N} + \frac{k_i}{N} \bar T \right) = \frac{T}{1 - 2 K/N}.
\end{equation}
The quantity $K$ is defined as $K = \sum_{j=1}^M k_j$. This result does not depend on the choice of destinations of passengers $\zeta_{ji}$. The terms disappear in the summation in Eq. (\ref{eq:that}) because $\sum_{i=1}^M \zeta_{ji} = 1$. The resulting average waiting time for passengers at bus stop $i$ is:
\begin{equation}
\text{W}^{\text{case 1}}_i = \frac{\bar T}{2} \left(1 - \frac{k_i}{N} \right) = \frac{T}{2} \left( \frac{N - k_i}{N - 2 K} \right).
\end{equation}
The average waiting time for the whole bus loop is the weighted average of the average waiting time at each bus stop:
\begin{equation}\label{eq:case1}
\text{W}^{\text{case 1}} = \frac{\sum_{i=1}^M k_i \text{W}^{\text{case 1}}_i}{K} = \frac{T}{2} \sum_{i=1}^M \left( \frac{k_i}{K} \left(\frac{N - k_i}{N - 2 K} \right)\right) = \frac{T}{2} \left( \frac{N K - \sum_{i=1}^M k_i^2}{N K - 2 K^2} \right).
\end{equation}

\subsection{Express buses} \label{subsec:2}
For this second case let us partition the set $\Omega$ of $N$ buses into disjoint subsets $\{\Omega_b \}_{b=1}^{N_S}$ ($\cup_b \Omega_b = \Omega$ and $\Omega_i \cap \Omega_j = \emptyset , \; \forall i \ne j$). Each subset of buses $\Omega_b$ boards passengers from a non-empty disjoint subset $\Theta_b$ of the set of bus stops in the loop $\Theta$ ($\cup_b \Theta_b = \Theta$ and $\Theta_i \cap \Theta_j = \emptyset , \; \forall i \ne j$) and let passengers alight at any bus stop in the loop.
This ensures that different subsets of buses do not interact with one other, in the sense that the total time spent on the road for each loop, $\bar T_b$, only depends on the number of buses in $\Omega_b$ and the ratio between passenger arrival rates and boarding/alighting rate of the bus stops in $\Theta_b$ (see Eq. (\ref{eq:that}) ). It can be shown from Eq. (\ref{eq:tau}) for buses in $\Omega_b$: by not boarding passengers from bus stops $\notin \Theta_b$, their effective $k_i$ is zero and the sum of the time spent dwelling at bus stops for boarding and alighting passengers only depends on $k_j$ for $j \in \Theta_b$ and the number of buses serving them $N_b = \vert \Omega_b \vert$. As in the previous case, the destination of passengers $\zeta_{ij}$ do not impact the average waiting time. 
We define the total time taken by buses $\in \Omega_b$ to complete a loop as $\bar T_b$. Buses in the same subset $\Omega_b$ will bunch and move as a single platoon but they are effectively decoupled from buses $\in \Omega_{b'} , b' \ne b$, so the different platoons do not bunch.

By considering a bus stop $i \in \Theta_b$ served by buses $\in \Omega_b$, the average waiting time for passengers at bus stop $i$ is given by Eq. (\ref{eq:wt}) with $\bar T = \bar T_b$.
This can be solved for bus stop $i$ as:
\begin{equation}\label{eq:wtstopexpress}
\text{W}^{\text{case 2}}_i =  \frac{T}{2} \left( \frac{N_b - k_i}{N_b - 2 K_b} \right),
\end{equation}
where $K_b$ is defined as $K_b = \sum_{j \in \Theta_b} k_j$. The average waiting time for the whole system is obtained by weighting the average waiting time at each bus stop by the number of passengers arriving at such bus stop, which is proportional to $k_i$.
\begin{equation}\label{eq:case2}
\text{W}^{\text{case 2}} = \frac{T}{2} \sum_{b=1}^{N_S} \sum_{i \in \Theta_b} \left( \frac{k_i}{K} \left( \frac{N_b - k_i}{N_b - 2 K_b} \right) \right) = 
\frac{T}{2}  \sum_{b=1}^{N_S}  \frac{K_b N_b - \sum_{i \in \Theta_b} k_i^2}{K N_b - 2 K K_b}.
\end{equation}
This result is not immediately comparable with regular buses Eq. (\ref{eq:case1}) because different partitions of buses and bus stops will give different results. Examples that compare case 1 and case 2 are discussed in sections \ref{sec:example} and \ref{sec:exampleNTU} and shown in Fig \ref{fig:kAkB}.

The general case where buses can serve arbitrary bus stops is not tractable with our method. In section \ref{sec:MARL} we introduce a computational tool to study this problem. One of the situations discovered by our algorithm is analysed in Ref. \cite{saw2020chaos} and reported in section \ref{sec:chaos}. The complex interaction between buses leads to chaotic behaviour.

\subsection{Scenario 1: a morning commute}\label{sec:example}
In this section we consider an example to compare the effect of different configurations of buses. The two cases of regular and express buses where no dynamic control is involved are examined. Consider a loop with three bus stops: $\text{A}$, $\text{B}$ and $\text{C}$ and two buses: $\text{X}$ and $\text{Y}$. Passengers arrive at bus stops A and B with rates $k_A = s_A/l$ and $k_B=s_B/l$ and everyone wants to alight at bus stop C, where no passengers board from there ($k_C = 0$), so $\zeta_{AC} = \zeta_{BC} = 1$. This can be interpreted as a simplified morning commute model where passengers leave their residential area (A and B) and travel to C: their workplace or a transportation hub such as a train station.

In this loop, we want to compare the average waiting time for passengers at bus stops A and B in the following cases:
\begin{enumerate}
\item Both buses serve all the bus stops as in case 1 in section \ref{sec:cases};
\item One bus serves bus stop A, the other bus serves B; both deliver passengers to C: ``express buses" as in case 2 in section \ref{sec:cases}.
\end {enumerate}
The first case is what normally happens in many bus loops where stop skipping is not implemented. The advantage of it is that the total load is evenly shared, speeding up the boarding and alighting time by parallelising it but the disadvantage is bunching. Uncontrolled buses in a loop inevitably bunch which is generally bad in terms of average waiting time. For the express buses, bunching is avoided because the stopping time of a bus is independent of the position of the other bus.

From Eqs. (\ref{eq:case1}) and (\ref{eq:case2}), the average waiting times are, respectively:
\begin{align}\label{eq:simplec1}
\text{W}^{\text{case 1}} &= \frac{T}{2} \left( \frac{k_A}{K} \left( \frac{2-k_A}{2 - 2 K} \right) + \; \frac{k_B}{K} \left( \frac{2-k_B}{2 - 2 K} \right) \right); \\
\label{eq:simplec2}
\text{W}^{\text{case 2}} &= \frac{T}{2} \left( \frac{k_A}{K} \left( \frac{1-k_A}{1 - 2 k_A} \right) + \; \frac{k_B}{K} \left( \frac{1-k_B}{1 - 2 k_B} \right) \right).
\end{align}

\begin{figure}[!ht]
\includegraphics[width=17cm]{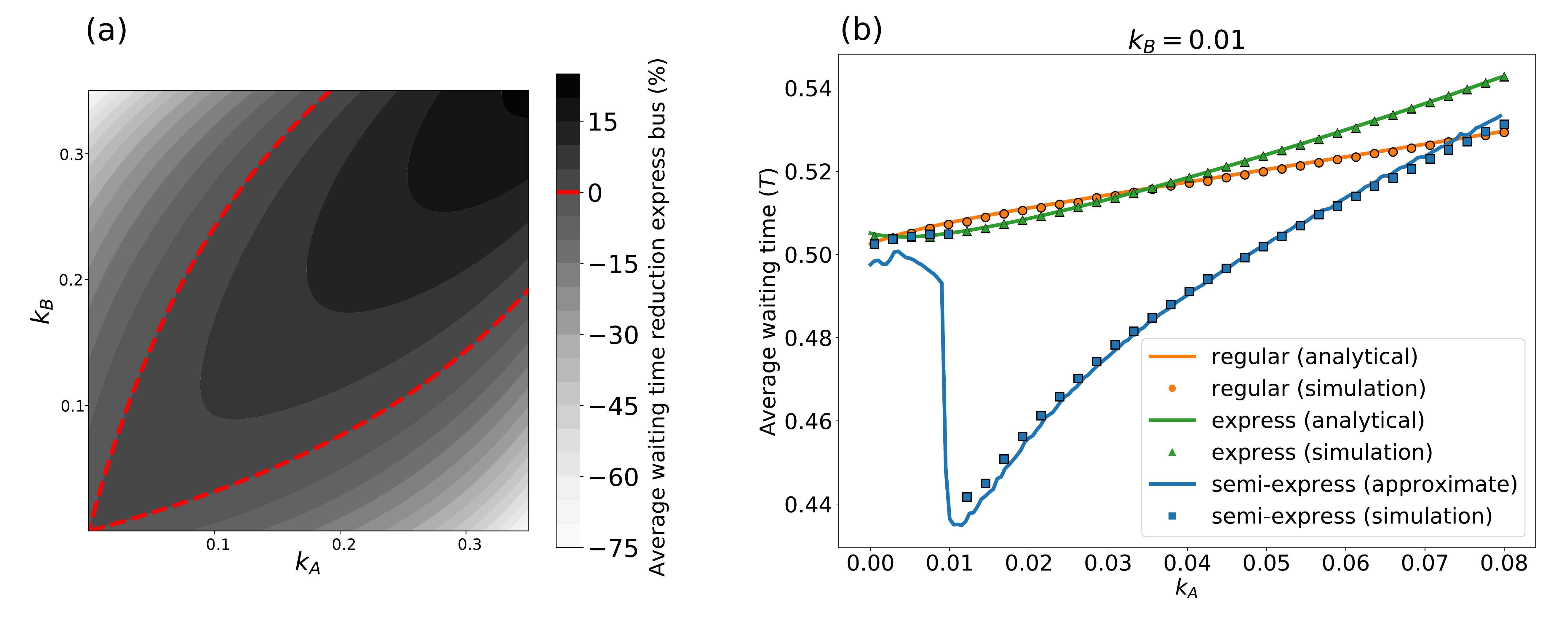}
\caption{(a) Average waiting time advantage of two express buses over regular buses as a function of $k_A$ and $k_B$  in the three-bus stop scenario examined in section \ref{sec:example}. It is defined as: $( \text{W}^{\text{case 1}} - \text{W}^{\text{case 2}}) / \text{W}^{\text{case 1}} \times 100$ (\%). The dashed red contour line represents the region where express and regular buses have the same average waiting time. For similar values of $k$, ($k_A \approx k_B$), express buses perform better than two regular buses serving all the bus stops in terms of average waiting time and the advantage increases with increasing demand, i.e. high $k$.\\
(b) Comparison of average waiting time (in units of $T$) analytically calculated from Eqs. (\ref{eq:case1}) and (\ref{eq:case2}) with the average waiting time from our simulation environment used to study the general case of bus assignment in section \ref{sec:MARL} for regular and express buses. The semi-express case is the one discovered by the reinforcement learning algorithm and it is discussed in section \ref{subsec:rl_ABC}. There is a sharp transition in performance for $k_A = k_B = 0.01$. In Ref. \cite{saw2020chaos} we show how to obtain an approximation for the average waiting time in the semi-express configuration (continuous blue line) and that this system is chaotic.}\label{fig:kAkB}
\end{figure}

It is not immediately transparent what the best configuration is, but it helps to consider two opposite edge cases first.
Suppose that $k_B=0$, in this situation, we expect the configuration in case 1 to perform better because one of the buses in case 2 is not doing anything since there are no passengers to board at bus stop B. This is indeed reflected by Eqs. (\ref{eq:simplec1}) and (\ref{eq:simplec2}): $(2 - k_A)/(2 - 2 k_A) < (1 - k_A)(1 - 2 k_A) , \; \forall \; k_A \in \left(0 , 1/2 \right)$. The other simple case to consider is $k_A = k_B = k$. In this situation, the load of passengers is distributed equally between the two express buses and the average waiting time is always lower for express buses in this configuration: $(2 - k)/(2 - 4 k) = (1 - k/2)/(1 - 2 k) > (1 - k)/(1 - 2 k) , \; \forall \; k \in \left(0 , 1/2 \right)$. Notice that the advantage of express buses grows for a very busy line (high $k$). Fig. \ref{fig:kAkB} shows a comparison of express and regular buses with varying $k_A$ and $k_B$.
This same scenario with $k_A = 0.015$ and $k_B = 0.010$ is examined in section \ref{subsec:rl_ABC} with a reinforcement learning approach and a novel bus strategy that sits between case 1 and case 2 emerges. Such a strategy is then reported in section \ref{sec:chaos} and more in details in Ref. \cite{saw2020chaos}. Fig. \ref{fig:kAkB} (b) compares it with regular and express buses for different values of $k_A$ with $k_B$ set at $0.01$.

\subsection{Scenarios 2, 3, 4: a university campus bus loop}\label{sec:exampleNTU}
Here we examine three empirically-motivated scenarios of a 5.5 km loop with 12 bus stops that surrounds the main campus of Nanyang Technological University, Singapore. 
\begin{enumerate}
\setcounter{enumi}{1}
    \item A simplified scenario in which passengers arrive at six origin bus stops with $k_i=0.0547, i=7, \cdots, 12$ and are directed to six destination bus stops $k_i=0, i=1, \cdots, 6$. The passengers alight at the opposite bus stop from where they boarded, i.e. for an origin bus stop ($i=7, \cdots, 12$), $\zeta_{ij} = 1$ for $j = 1 + {(i+5)}\bmod{12}$, zero otherwise. Six buses are present in the loop. A variation of this with random $\zeta_{ij}$ is also studied in Ref. \cite{saw2021vae}.
    \item A realistic ``lull" scenario (weekday afternoon) with values of $k_i$ as measured in Ref. \cite{saw2019no} and three buses. Passengers alight uniformly, i.e. $\zeta_{ij} = 1/11 , \; \forall j \neq i$, zero otherwise.
    \item A realistic ``busy" scenario (weekday morning peak hour) with values of $k_i$ as measured in Ref. \cite{saw2019no} and six buses. Passengers alight uniformly, i.e. $\zeta_{ij} = 1/11 , \; \forall j \neq i$, zero otherwise.
\end{enumerate}
For situations 3 and 4, the values of $k_i$ are estimated from the real-time positional data of the buses measured between 16 and 20 April 2018 \cite{saw2019no}.
For scenario 2, the values of $k_i$ sum to the same value as in the ``busy" scenario 4, which means that the total number of passengers arriving per unit time in the whole system is the same in both configurations.

In the three scenarios in this section, it is possible to computationally enumerate all the different ways of assigning express buses, as defined in section \ref{sec:cases}, to bus stops and compute the average waiting time with Eq. (\ref{eq:case2}) to find the best setting.

For the simplified scenario (2), the best average waiting time is achieved by assigning each of the six buses to a different one of the six origin ($k_i > 0$) bus stops, effectively subdividing the passengers equally to each bus. The reduction in average waiting time over regular buses is approximately $4.6\%$.

Table \ref{Table_nut_lull} reports the optimal distribution of the 12 bus stops among the 3 express buses of scenario 3. The total number of passengers served by every bus is approximately the same: $\sum_{i \in \Omega_{B_1}} k_i \approx \sum_{i \in \Omega_{B_2}} k_i \approx \sum_{i \in \Omega_{B_3}} k_i$.
The reduction of the average waiting time of express buses is around $1.7\%$. 

For the ``busy" period (scenario 4), the optimal distribution of the bus stops is shown in Table \ref{Table_nut_busy}. As in the previous scenario, the buses are distributed in such a way that the number of passengers is as uniformly distributed as possible. The advantage, compared with regular buses, is a reduction of $3.6\%$ of the average waiting time.

Much better performance can be achieved by lifting the constraints of express buses and allow for arbitrary assignments of buses to bus stops. Unfortunately, our exact analytical theory cannot deal with such cases so a reinforcement learning algorithm is introduced in section \ref{sec:MARL} to address this general case. The framework applied to the scenarios listed above (section \ref{subsec:rl_ntu_busy}) shows reductions of up to $32\%$ in average waiting time compared with regular and express buses. See Fig. \ref{fig:performance_comparison_rel} for a summary of the performance achieved in different scenarios.

\begin{table}[h]
\begin{tabular}{|l|l|l|l|l|l|l|}
\hline
\textbf{Bus stop} & H4 & IC & SPMS & WKW & CEE & LWN  \\
\hline
$\boldsymbol{k_i}$ & 0.001 & 0.023 & 0.015 & 0.005 & 0.016 & 0.040 \\
\hline
\textbf{Bus assigned} & $B_1$ & $B_1$ & $B_1$ & $B_2$ & $B_3$ & $B_2$ \\
\hline \hline
\textbf{Bus stop} & H3 & H14 & CH & H10 & H8 & H2  \\
\hline
$\boldsymbol{k_i}$ & 0.018 & 0.035 & 0.024 & 0.030 & 0.07 & 0.010 \\
\hline
\textbf{Bus assigned} & $B_1$ & $B_3$ & $B_3$ & $B_2$ & $B_1$ & $B_1$ \\
\hline
\end{tabular}
\caption{Values of $s_i/l = k_i$ for the bus stops in the Nanyang Technological University campus bus loop as measured in Ref. \cite{saw2019no}. Express buses $B_1$, $B_2$ and $B_3$ are assigned to optimise the average waiting time according to Eq. (\ref{eq:case2}).}
\label{Table_nut_lull}
\end{table}

\begin{table}[h]
\begin{tabular}{|l|l|l|l|l|l|l|}
\hline
\textbf{Bus stop} & H4 & IC & SPMS & WKW & CEE & LWN  \\
\hline
$\boldsymbol{k_i}$ & 0.000 & 0.063 & 0.026 & 0.033 & 0.008 & 0.027 \\
\hline
\textbf{Bus assigned} &  & $B_1$ & $B_2$ & $B_3$ & $B_3$ & $B_2$ \\
\hline \hline
\textbf{Bus stop} & H3 & H14 & CH & H10 & H8 & H2  \\
\hline
$\boldsymbol{k_i}$ & 0.067 & 0.001 & 0.006 & 0.063 & 0.003 & 0.031 \\
\hline
\textbf{Bus assigned} & $B_4$ & $B_5$ & $B_5$ & $B_6$ & $B_5$ & $B_5$ \\
\hline
\end{tabular}
\caption{Values of $s_i/l = k_i$ for the bus stops in the Nanyang Technological University campus bus loop as measured in Ref. \cite{saw2019no}. In that paper $k_{\text{H4}} < 0$ because of noise in the measurement, here it is set to $0$ hence H4 acts only as a destination stop: $\tau_{\text{H4}}^{\text{board}} = 0$ so no bus is assigned to board passengers from there. Express buses $B_1$, $B_2$, $B_3$, $B_4$, $B_5$ and $B_6$ are assigned to optimise the average waiting time according to Eq. (\ref{eq:case2}).}
\label{Table_nut_busy}
\end{table}

\subsection{Insights}\label{sec:insight}
The average waiting times for case 1 (regular) and case 2 (express) do not depend on the destination of passengers. This means that adding a destination bus stop, i.e. a bus stop with $k_i=0$ where passengers can only alight does not change the average waiting time. What matters is how many passengers board ($k_i > 0$) and the available buses ($N$ or $N_b$).
The total time for all of the passengers to eventually alight is the same, regardless of where they alight.
In a fully symmetric situation with $k_i = k = K/M, \forall i$ and $N = M$, the average waiting time is always lower with express buses. From Eqs. (\ref{eq:case1}) and (\ref{eq:case2}): $(1 - k/N)/(1 - 2k) > (1 - k)/(1 - 2 k)$, $\forall N > 1$ and $k < 1/2$.

Splitting a bus stop $i$ into two non-empty bus stops $i_1$ and $i_2$ with $k_{i_1} + k_{i_2} = k_i$ always increases the average waiting time when buses bunch. This can be seen from the numerator of the corresponding term in Eq. (\ref{eq:case1}) where:
\begin{align*}
k_i (N - k_i) =&  (k_{i_1} + k_{i_2}) N - (k_{i_1} + k_{i_2})^2 < \\
k_{i_1} (N - k_{i_1})+k_{i_2} (N - k_{i_2}) =&  (k_{i_1} + k_{i_2}) N - (k_{i_1}^2 + k_{i_2}^2) ,
\end{align*}
since $(k_{i_1} + k_{i_2})^2 > k_{i_1}^2 + k_{i_2}^2$.

The results in Eqs. (\ref{eq:case1}) and (\ref{eq:case2}) show that there is a limit to the number of passengers that can be served in a bus line. For regular buses, this condition is $2K < N$ while for a subset $\Omega_b$ of express buses serving a subset $\Theta_b$ of bus stops the constraint is $2 K_b < N_b$. It's easier to see this in terms of arrival rates $s_i$ and boarding rate $l$. The necessary condition becomes $2 \sum_i s_i < N l$ for regular buses.
If the inequality is not satisfied, buses cannot keep up with the demand of passengers. At every time step $\sum_i s_i$ passengers arrive: they have to board a bus and be delivered to their destination (alight). The $N$ buses collectively can only board or alight up to $N l$ passengers per unit time.

With the framework introduced we can calculate the best assignment of $N$ express buses to $M$ bus stops to minimise the average waiting time. The intuition suggests that more buses should serve crowded bus stops (high $k_i$). Let us consider the simple example of two bus stops with $k_A$ and $k_B$. The $N$ buses are divided in $N_A$ that board from bus stop A and $N - N_A$ from bus stop B. Eq. (\ref{eq:case2}) becomes:
\begin{equation}\label{eq:distributeN}
\text{W}^{\text{case 2}} = \frac{T}{2 (k_A + k_B)} \left( k_A \left( \frac{N_A - k_A}{N_A - 2 k_A} \right) + k_B \left( \frac{N - N_A - k_B}{N - N_A - 2 k_B} \right) \right).
\end{equation}
The value of $N_A$ that minimises Eq. (\ref{eq:distributeN}) is $N^*_A = N k_A /K$ so the ideal distribution of express buses is proportional to the values of $k_i$ of the bus stop served for fixed $K = k_A + k_B$.
The complementary example is the problem of assigning two express buses $B_1$ and $B_2$, to $M$ bus stops, each with $k_i=k$. The first bus $B_1$ serves $M_1$ bus stops, while $B_2$ serves the remaining $M-M_1$. Following Eq. (\ref{eq:case2}):
\begin{equation}\label{eq:distributeM}
\text{W}^{\text{case 2}} = \frac{T}{2 M} \left(1 -  k \right) \left( \frac{M_1}{1 - 2 M_1 k} +\frac{M - M_1}{1 - 2 \left( M - M_1\right) k} \right).
\end{equation}
The value of $M_1$ that minimises the equation above is $M/2$ so the average waiting time is minimised when both buses board the same number of passengers, which is in line with what is found by enumerating all the possible combinations for empirical $k_i$ in section \ref{sec:exampleNTU}.

\begin{section}{A reinforcement learning platform for a bus loop}\label{sec:MARL}
The analytical formalism discussed in section \ref{sec:cases} is limited by our inability to treat exactly the interaction that can arise if there is an intermediate situation between case 1 (regular) and case 2 (express) of section \ref{sec:cases}. Such situations can involve buses boarding passengers from intersecting subsets of the bus stops $\Omega_a \cap \Omega_b \neq \emptyset$, which means that there is at least one bus stop where the time taken to board a bus $\tau_i^{\text{board}}$ depends on when another bus left that bus stop. Section \ref{sec:chaos} examines an example of such situations in a simplified setting showing that the interaction between buses can lead to chaotic behaviour.

The objective of this section is to introduce a platform to find the best configuration for a given bus loop specified by the number of buses, $N$, the arrival rates over boarding/alighting rate of the $M$ bus stops, $\{k_i\}_{i=1}^M$, and the probabilities $\{\{\zeta_{ij} \}_{i=1}^M \}_{j=1}^M$ for a passenger in $i$ to be directed to $j$.
The framework of choice is tabular Q-learning \cite{watkins1989learning} \cite{watkins1992q}.

This algorithm runs on a bus simulation. More details are available in Appendix \ref{appendix:RL}. Fig. \ref{fig:kAkB} (b) shows a comparison of regular and express buses between the analytical formulas Eqs. (\ref{eq:case1}), (\ref{eq:case2}) and the simulation for the scenario 1 in section \ref{sec:example}.

\begin{subsection}{Reinforcement learning}\label{subsec:RL}
The bus problem can be framed as a Markov Decision Process (MDP) \cite{sutton2011reinforcement}.
A MDP is defined as a tuple $(S, A_s, P_a, c_a)$ where:
\begin{itemize}  
\item $S$ is the discrete set of the possible states $s$ of the system;
\item $A_s$ is the set of actions available in a state $s$;
\item $P(s' \mid s, a)$ is the probability of transitioning from state $s$ to $s'$ after performing the action $a$ and satisfies the Markov property, i.e. it depends only on the current state $s$ and action $a$;
\item $c_a$ is the cost incurred for the transition from $s$ to $s'$ after performing the action $a$.
\end{itemize}
The goal of reinforcement learning is to find an optimal policy for an agent (a bus in this case) which is a set of rules for picking an action $a$ based on the state $s$ that minimises the expected discounted cost $C$. This quantity is defined as the sum of the costs discounted by a factor $\gamma \in [0, 1]$ for every future state visited from $s = s(t=0)$: $C(s) = \sum_{t=0}^{\infty} c_t \gamma^t$. In the formula $c_t$ is the cost incurred transitioning from state $s(t)$ to $s(t+1)$ via the action $a(t)$. In general, the state of the system cannot be fully observed by the agents: this can be for technological reasons (some quantities of interest may not be practically measurable) and for simplification reasons (some information are not needed for making a decision). Such variations are called Partially Observable MDPs  \cite{Astrom1965}. 

The algorithm of Q-learning \cite{sutton2011reinforcement} \cite{watkins1989learning} can deal with unknown probability of transition from state $s$ to $s'$ and a partially observed state. An implicit model of the environment as transition probabilities is obtained by exploring the possible state-action pairs and averaging all the past experiences following Eq. (\ref{eq:watkins1989learningearning}).
The estimator of the expected discounted cost of a state-action pair, $Q(s, a)$, is learned by a trial-and-error process in a simulation, observing the cost incurred and guessing the future discounted cost in the next state $s'$ using $Q(s', a')$ itself. The optimal policy is obtained by taking the action $a$ that gives the lowest expected discounted cost $C$ in a given state $s$ so $a = \operatorname{arg\,min}_{a_i} Q(s, a_i)$ where $Q(s,a)$ is learned by the agent while interacting with the environment.
The update rule (training) of $Q(s, a)$ after observing the cost $c_a$ obtained and the next state $s'$ is described by the following iterative equation \cite{watkins1989learning}:
\begin{equation}\label{eq:watkins1989learningearning}
Q^{\operatorname{new}}(s, a) = (1-\alpha) Q(s, a) + \alpha \left(c_a + \gamma  \operatorname{arg\,min}_{a'} Q(s', a')  \right) .
\end{equation}
The parameter $\alpha \in [0, 1]$ is called learning rate and it is the weight of the most recent information in computing the estimated return.

For this application, each agent (bus) has a separate policy, each action is taken when a bus reaches a bus stop, after the passengers alight (if any). The possible actions for a bus at a bus stop are: allow passengers to board (stay) or leave the bus stop without boarding any passenger (skip). The state observed, which is the information used to decide an action, is the current bus stop number.
For a bus stop with $k_i = 0$, such as bus stop C in the example in sections \ref{sec:example} and \ref{subsec:rl_ABC}, only one action is allowed since there are no passengers to board.
The information about the bus stop number allows different actions for different bus stops, such as what happens with the express buses analysed in section \ref{sec:cases}.
The final goal is for the system to converge to a set of optimal rules for each bus that dictates which bus stops are served. This platform will be able to assign buses to bus stops in a general manner to minimise the cost selected, overcoming the limitations of the analytical approach described in section \ref{sec:cases}.

In this particular application, having more than one agent operating on the same environment, the Markov property of $P(s' \mid s, a)$ is not satisfied because the actions taken by other agents influence the actual state of each agent. Q-learning can still work in this case \cite{Busoniu2008} but without the Markov property, there is no theoretical guarantee on the convergence and uniqueness of the optimal policy \cite{watkins1989learning}. In the more complicated scenarios in sections \ref{subsec:rl_simplified}, \ref{subsec:rl_ntu_lull} and \ref{subsec:rl_ntu_busy} we observe that the policy does not always converge to an optimal solution and there are multiple policies with very similar performance. 

This framework finds solutions to the problem of assigning buses to bus stops in a loop by the dynamics of reinforcement learning. The configuration of buses can then be implemented without any dynamical control. 
More details about the algorithm and the parameters used are in appendix \ref{appendix:RL}.
\end{subsection}

\begin{subsection}{Costs to optimise}\label{sec:cost}
The objective of reinforcement learning is to find a policy that minimises a given quantity: the expected discounted cost $C$. Many choices are possible for the cost, for example, the travelling time \cite{Osuna1972}, the adherence to a schedule  \cite{Chen2015} \cite{Xuan2011} or the deviation from a staggered position \cite{saw2019intelligent} \cite{Moreira-Matias2016}. Here we consider the average waiting time at bus stops, as measured by the buses: $c = \text{WT}$. It is computed by averaging the time waited at their bus stops by all the passengers waiting for a bus at a given time.


In this paper, we calculate $c_a$ at the midpoint between the bus stop (state $s$) where the action $a$ is taken and the following bus stop (state $s'$). In appendix \ref{appendix:RL} we mention other possible choices and why this solution works best for our bus system.
\end{subsection}

\begin{subsection}{Scenario 1 revisited: a morning commute}\label{subsec:rl_ABC}
As a first example, we examine the same scenario that was investigated analytically in section \ref{sec:example}. The loop comprises two buses, X and Y, and three bus stops, A, B and C with $k_A, k_B > 0$ and $k_C = 0$. All the passengers want to alight at bus stop C.

It has been shown in Fig \ref{fig:kAkB} (a) that, depending on how different $k_A$ and $k_B$ are, express buses, in the form of one bus boarding passengers at A and the other boarding passengers at B, can lead to a lower average waiting time if $k_A \approx k_B$ or longer average waiting time if the arrival rates are very inhomogeneous, compared with regular buses.
The motivation for the use of reinforcement learning is to validate the system over the exact results obtained but, more importantly, to explore if different strategies can emerge. 

We present the setting of $k_A = 0.015$ and $k_B = 0.010$ which sits as an intermediate situation where express and regular buses perform very similarly.
The analytical prediction for the average waiting time in the case of regular and express buses are respectively approximately $0.509$ and $0.507$ in units of $T$, so express buses reduce the average waiting time by less than $1\%$ in this specific example.

The best strategy found is unexpected: one bus boards both A and B and one bus boards only passengers at bus stop B. We refer to this configuration as \emph{semi-express}. This solution significantly outperforms both express and regular buses in terms of average waiting time at approximately $0.446 T$. The performance of this configuration is also reported in Fig. \ref{fig:kAkB} (b), in section \ref{sec:chaos} and in more details in Ref. \cite{saw2020chaos}.

It is worth noting that bus stop A, despite being the more crowded one, is counterintuitively served by only one bus. This effectively makes such bus slower in completing a loop. It has been shown \cite{saw2019} that regular buses with different natural periods can unbunch and bunch repeatedly, in contrast with regular buses with the same natural period that once bunched will stay bunched indefinitely. What happens with the semi-express configuration presented is somewhat similar to the situation in Ref. \cite{saw2019}, effectively eliminating the risk of getting stuck in a bunched configuration by slowing down one of the buses more than the other. By comparison, the alternative semi-express configuration, called \emph{alt-semi-express}, with X boarding both A and B and Y boarding only A, gets trapped in the bunched state for the parameters $k_A$ and $k_B$ chosen. The reason for this behaviour is that the extra time spent boarding at bus stop B by bus X is not enough to cause unbunching, so X will initially be left behind by Y, but it will reach Y at bus stop A before Y finishes boarding, hence bunching again. This solution has worse performance in terms of average waiting time compared with the semi-express configuration discovered by the algorithm for $k_A > k_B$. The average waiting time is approximately the same as regular and express buses for the chosen parameters $k_A = 0.015$ and $k_B = 0.010$. We found the alt-semi-express configuration to be optimal in the regime $k_A < k_B$ while semi-express buses cannot unbunch in this scenario, leading to longer average waiting time.

\end{subsection}

\begin{subsection}{Scenario 2 revisited: an empirically-based bus loop with six buses serving six origin and six destination bus stops}\label{subsec:rl_simplified}
In this application, we study the reinforcement-learning method applied to scenario 2 in section \ref{sec:exampleNTU}.
The scenario is a simplified model of a situation where commuters leave their residence place ($i = 7, 8, \cdots 12$ origin bus stops with $k_i = 0.0547$) to go to their working/studying place ($i = 1, 2, \cdots 6$ destination bus stops with $k_i = 0$).

Due to the symmetry of the system and the presence of multiple agents, there are several equivalent solutions.
Different runs of the algorithm lead to different results but very similar performance.

The reductions in average waiting time compared with regular buses is typically (median) approximately $26\%$ and up to $32\%$ in the best run out of 12 trials. The median and best performance are presented in Fig. \ref{fig:performance_comparison_rel}. The best configuration of Q-tables found is reported in Fig. \ref{fig:qTable_s_wt}.

\begin{figure}[!ht]
\includegraphics[width=16cm]{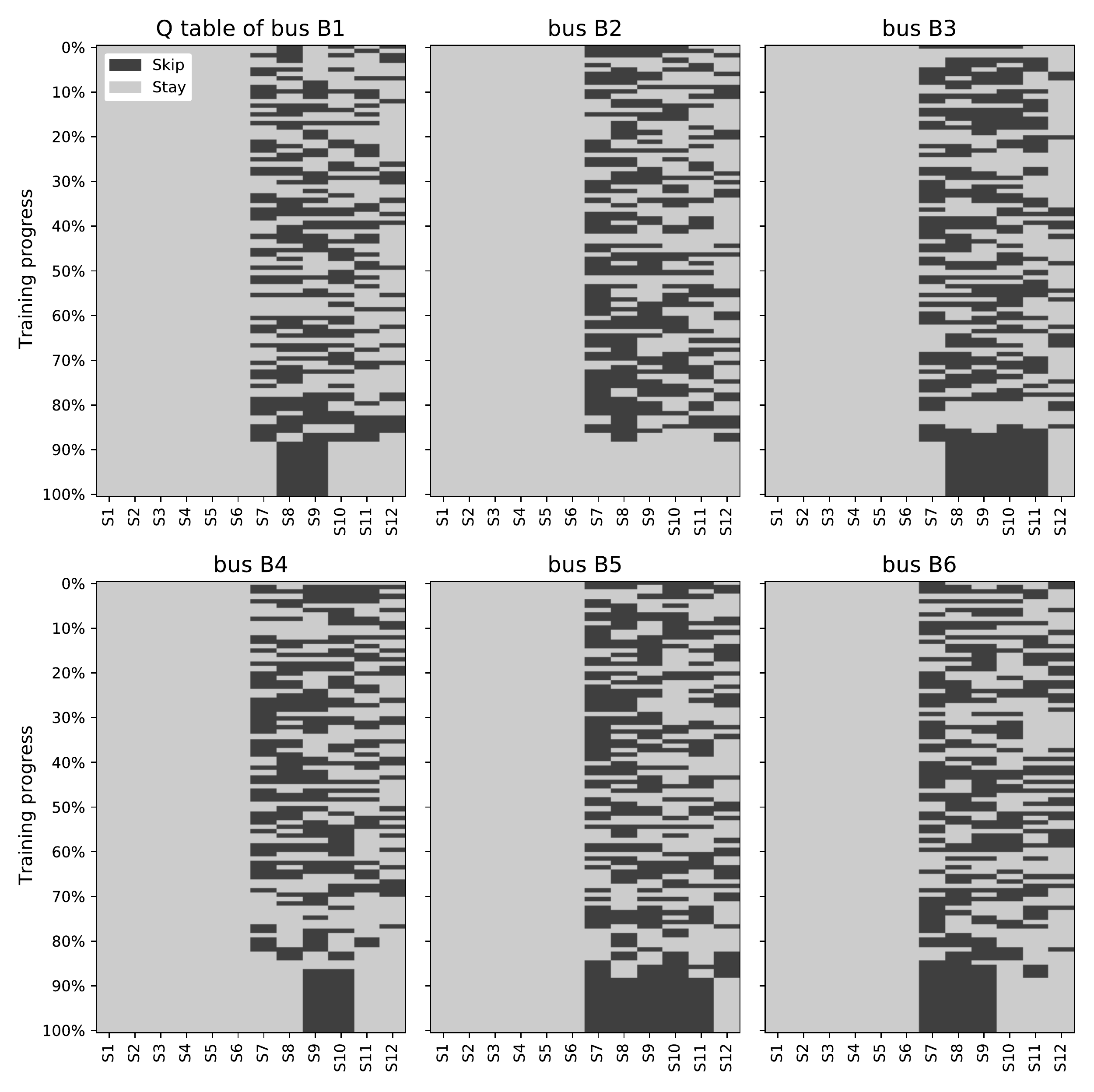}
\caption{Evolution of the Q-tables for the six buses in scenario 2, as described in section \ref{subsec:rl_simplified}. This represents the best solution found in terms of average waiting time at $0.376 T$ by the reinforcement learning algorithm. The possible actions are either to board from a bus stop (in light grey) or to not board from it (in dark grey). The decision can be made for origin bus stops ($k_i > 0$), so only at $S_6, S_7, \cdots S_{12}$ in this scenario. Buses always stop at the first six bus stops (destinations) to let passengers alight (if any).}\label{fig:qTable_s_wt}
\end{figure}


\end{subsection}

\begin{subsection}{Scenario 3 revisited: an empirically-based bus loop with three buses serving twelve bus stops}\label{subsec:rl_ntu_lull}
This section aims to apply the reinforcement learning framework introduced to the real-world problem of a university campus bus loop during off-peak hours. This is the same situation studied for express buses in section \ref{sec:exampleNTU} (scenario 3) and the values of $k_i$ used, taken from Ref. \cite{saw2019no}, are reported in Table \ref{Table_nut_lull}.

The algorithm converges to many different solutions. The median performance, over 12 runs, is $0.503 T$. The evolution of the Q-table of the best run is reported in Fig. \ref{fig:Q-lull} and the reduction in average waiting time is $14\%$ and $19\%$ compared with regular buses for the median and best cases respectively.
A summary of the performance, which includes express buses, is reported in Fig. \ref{fig:performance_comparison_rel}.

\begin{figure}[!ht]
\includegraphics[width=16cm]{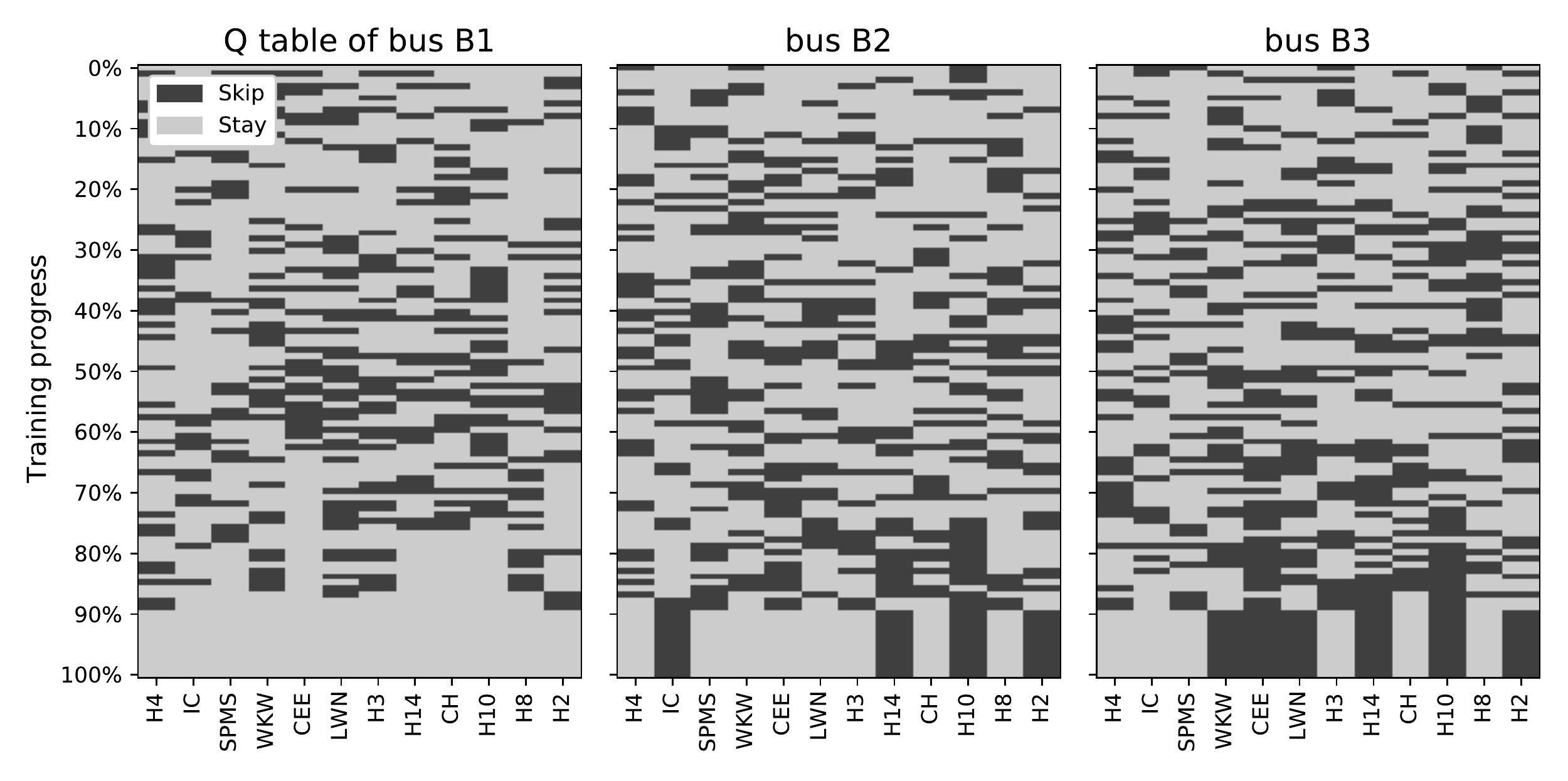}
\caption{Evolution of the Q-tables for the three buses in scenario 3, as described in section \ref{subsec:rl_ntu_lull}. This represents the best solution found in terms of average waiting time at $0.473 T$ by the reinforcement learning algorithm.}\label{fig:Q-lull}
\end{figure}

\end{subsection}

\begin{subsection}{Scenario 4 revisited: an empirically-based bus loop with six buses serving twelve bus stops}\label{subsec:rl_ntu_busy}
The last example proposed examines the fourth and last scenario of section \ref{sec:exampleNTU}, the challenging peak-hour situation in a university campus.
The number of possible ways of assigning $N$ buses to $M_k$ bus stops with $k_i>0$ is 
$A = (2^{M_k})^N$. 
The value for $A$ is computed considering that each of the $N$ buses can independently take two actions at each of the $M_k$ bus stops.
Since $k_{\text{H4}}=0$, $M_k=11$ in this setting, so $A \approx 10^{20}$, which is $10^9$ times the possible combinations for scenarios 2 and 3. As in the previous scenarios, different runs lead to different outcomes. 
The median and best performance in terms of average waiting time over 12 runs are respectively $0.441 T$ and $0.421 T$, reducing the average waiting time by up to $24\%$ compared to regular buses. The evolution of Q-tables of the six buses for the best-performing case is reported in Fig \ref{fig:Q-busy} and a comparison of the average waiting time with different configurations is in Fig. \ref{fig:performance_comparison_rel}.

\begin{figure}[!ht]
\includegraphics[width=16cm]{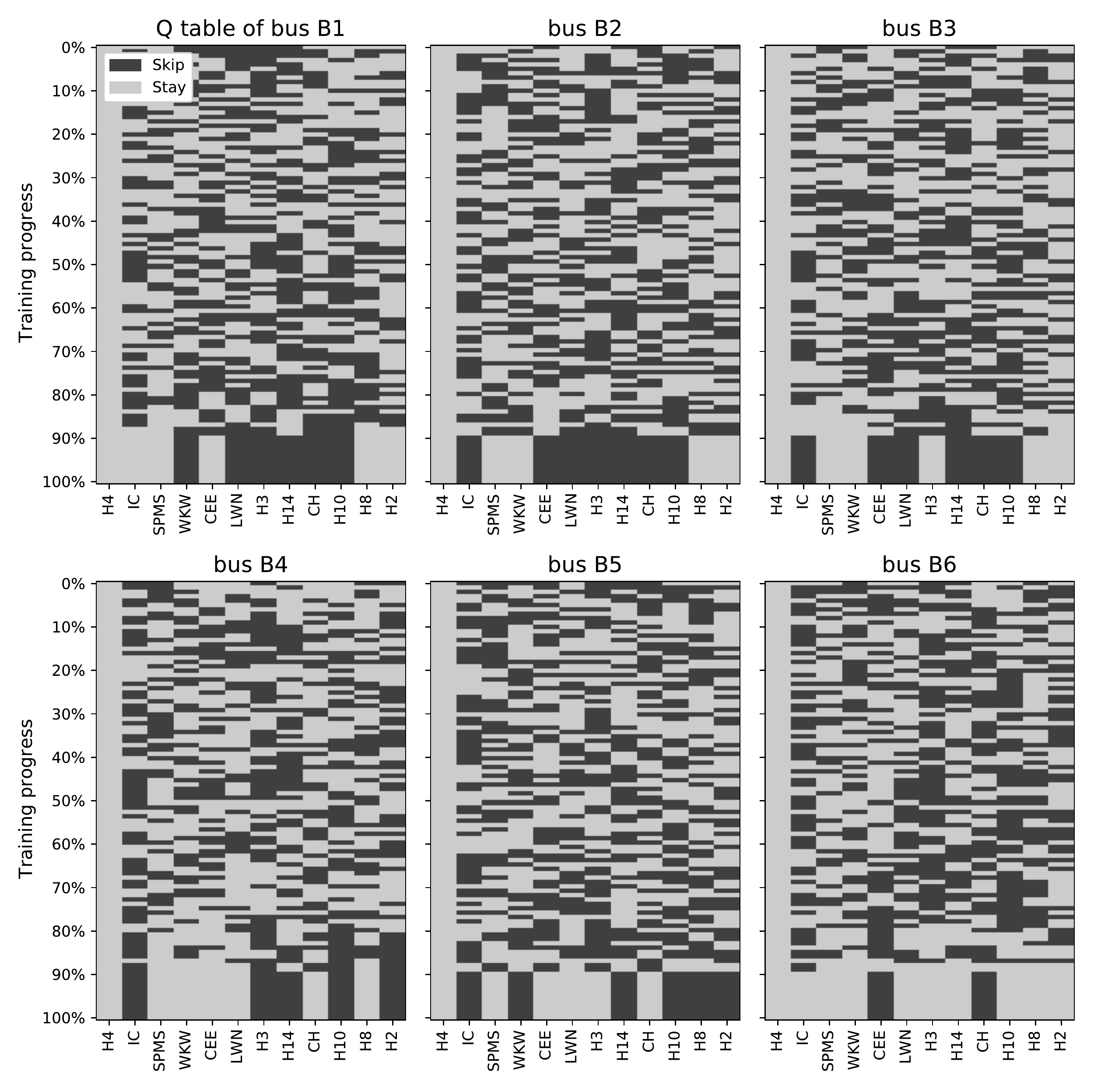}
\caption{Evolution of the Q-tables for the six buses in scenario 4, as described in section \ref{subsec:rl_ntu_busy}. This represents the best solution found in terms of average waiting time at $0.421 T$ by the reinforcement learning algorithm.}\label{fig:Q-busy}
\end{figure}

\end{subsection}


\begin{figure}[!ht]
\includegraphics[width=16cm]{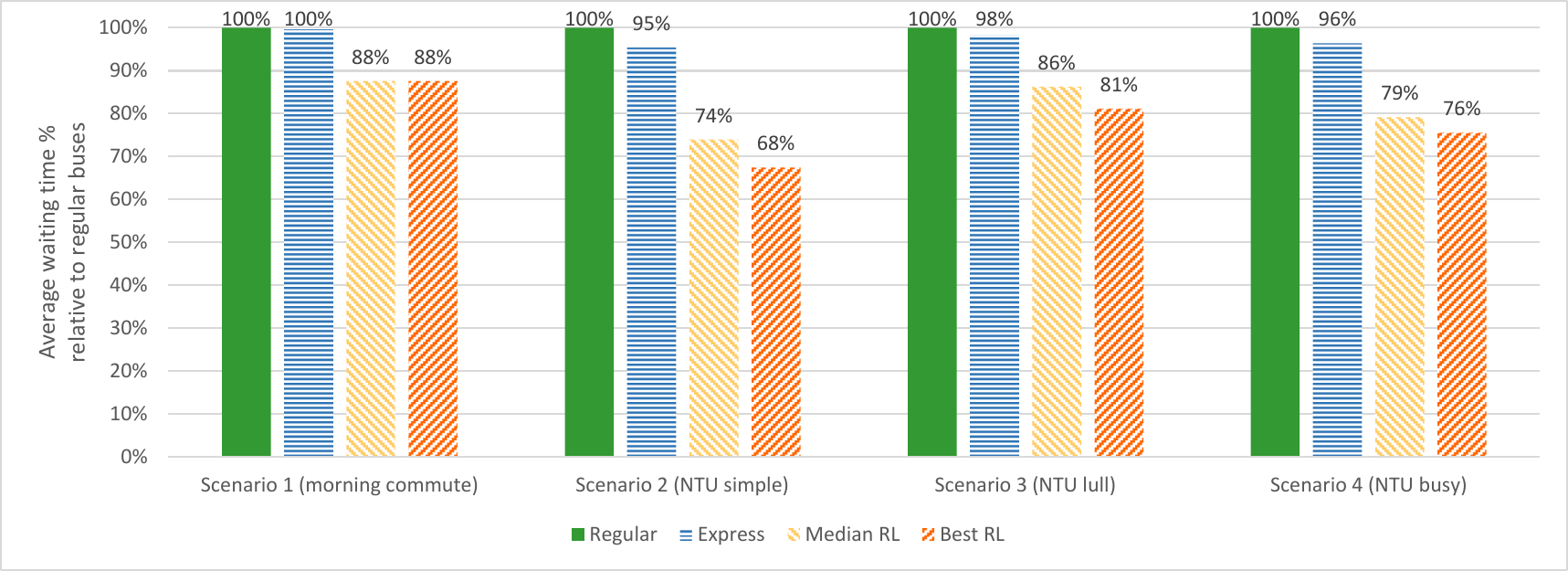}
\caption{Summary of the performance in terms of average waiting time (in percentage compared with regular buses as calculated in section \ref{sec:cases}) of regular and express buses with the median and best results obtained by the algorithm described in section \ref{sec:MARL} for the four scenarios examined in sections \ref{subsec:rl_ABC}, \ref{subsec:rl_simplified},  \ref{subsec:rl_ntu_lull} and \ref{subsec:rl_ntu_busy}. Appendix \ref{appendix:RL} contains more details on the algorithm and the simulation used.}\label{fig:performance_comparison_rel}
\end{figure}

\begin{subsection}{Summary of the algorithmic approach}
The problem of assigning $N$ buses to $M$ bus stops is complex due to the interaction between buses at bus stops. We can describe analytically the system for two opposite cases: every bus interacts with every other bus (regular buses) and disjoint independent groups of buses (express buses).
Reinforcement learning enables the exploration of any general interaction of buses and can optimise the average waiting time. Despite the theoretical limitations of applying Q-learning in a multi-agent setting, we were able to consistently reduce the average waiting time as compared with regular and express buses, in some cases by as much as $32\%$. This simple tool is scalable and generalisable to real-world bus systems, opening the possibility for practical applications.
\end{subsection}
\end{section}

\section{Chaotic behaviour in a bus loop}\label{sec:chaos}
The analytical results for regular and express buses presented in section \ref{sec:cases} are possible because there is no interaction between different groups of buses. In this section, we briefly present a case in which the buses interact at one bus stop. A more detailed analysis can be found in Ref. \cite{saw2020chaos}. This serves both as a justification for the use of reinforcement learning, given the difficulty to solve analytically for the average waiting time in general, and as a model where chaos arises from the interaction of buses in a simple loop. The system studied is the semi-express bus strategy, discovered by the reinforcement learning algorithm in the morning commute scenario in section \ref{subsec:rl_ABC} when the average waiting time is optimised.

In this situation, analogous to the one examined in section \ref{sec:example}, passengers from two bus stops, A and B, are served by two buses X and Y and delivered at bus stop C, the destination. Bus X is a regular bus and picks up passengers from both bus stops A and B; bus Y only picks up passengers at B. Both buses stop at C for the passengers to alight.

To study this case, two approaches are used. First, an approximate analytical map is introduced. The simplifying assumption used is to consider a fixed order of events, i.e. after bus X reaches bus stop B, bus Y reaches bus stop C, etc., every time. More details on the exact event list and the 10-d map used are in Ref. \cite{saw2020chaos}.
The approximation is not necessarily true but such formulation allows for analytical calculations of quantities such as the average waiting time and the dwelling time, showing chaotic behaviour. More importantly, it is possible to calculate the Liapunov exponents of the system, whose highest value is proven to be always positive. Despite the approximations made, the prediction for the average waiting time in the regime $k_A > k_B$ are remarkably accurate, as shown in Fig. \ref{fig:kAkB} (b).

The second approach involves a brute-force enumeration of all the possible states of the bus loop, i.e. all the possible positions of the buses with respect to one other and the bus stops. With this setting, we can study the dynamics of the system exactly in a simulation. The quantities of interest, such as the waiting time and the distance between buses, show chaotic behaviour (more details in Ref. \cite{saw2020chaos}). This further supports the argument for the emergence of chaos in the system.

The extensive enumeration of states and the discrete map method are built on a given policy and are useful to understand the dynamics and properties of a system. The implications of chaos in the real world are important since chaotic systems are inherently unpredictable, posing yet another challenge in designing reliable bus systems. While the discrete map is extendable to more complex scenarios, the enumeration approach is not scalable to bigger systems. The framework presented in section \ref{sec:MARL} can operate with an arbitrary number of buses and bus stops with the objective of finding a set of policies for the buses to optimise the average waiting time. 

\section{Conclusion}\label{sec:conclusion}
In the first part, we introduce the concept of express buses. The analytical results offer a way to compare express and regular buses in terms of average waiting time. The formula can also recommend how to assign express buses to subsets of bus stops as a function of the arrival rates $s_i = k_i l$.
Other insights can be drawn from the mathematical description of the problem: the average waiting time does not depend on the destination of the passengers, there is a minimum required number of buses as a function of the crowdedness of the bus loop and splitting a bus stop in two less crowded bus stops increases the average waiting time for regular buses.

The reinforcement learning platform introduced in section \ref{sec:MARL} can be applied to arbitrary models of bus loops with multiple heterogeneous bus stops and multiple buses, complementing the exact mathematical solutions for express and regular buses. The result obtained is a recommended policy where individual buses are assigned to serve bus stops in a configuration that minimises the average waiting time of passengers. In principle, it is also possible to optimise arbitrary costs. This method offers greater flexibility and it can potentially find applications in real-world bus systems.

In the final part, we report the configuration of semi-express buses emerged from the reinforcement learning framework. Chaotic behaviour is observed in the average waiting time and other relevant quantities.\cite{saw2020chaos}.

\section{Appendix:}
\appendix

\section{Description of the algorithm used}\label{appendix:RL}
The framework of choice for this particular application is tabular Q-learning \cite{sutton2011reinforcement} \cite{watkins1989learning} where each bus has an individual Q-table. When arriving at a bus stop, two actions can be taken: allow boarding or do not allow boarding. Alighting is always permitted.
States are represented by the current bus stop.

To explore the state and action space, $\epsilon\text{-greedy}$ is used: with probability $\epsilon$ a random action is taken, otherwise the action $a = \operatorname{arg\,min}_{a'} Q(s, a')$ is performed. The parameter $\epsilon$ is initialised at $0.2$ and exponentially decreased to $0.001$ when 90\% of the training episodes are completed. After that, $\epsilon$ is set to zero.
Each episode is $150 T$ long. Many revolutions are needed for the training since the effect of the actions taken may be more strongly reflected after several further actions in future. The system is trained for $30000$ episodes.

The other parameters used in this algorithm are the discount factor $\gamma$ that enters in the discounted cost $C = \sum_{t=0}^\infty \gamma^t c_t$ and the learning rate $\alpha$ that weights the importance of the new information observed over what has been learned so far in the update process that leads to the estimation of the discounted cost as a Q function. In this work, $\gamma = 0.5^{1/(4M)}$ where $M$ is the number of bus stops so the cost observed 4 revolutions in the past is discounted by a half compared to the current cost. The learning rate is set as $\alpha = 0.01$, except in the last $1\%$ of the episodes where $\alpha = 0$ to stop the learning and check the performance achieved.

We also use $n$-step Q-learning \cite{sutton2011reinforcement} with $n=6$ to reduce bias in the estimation of the expected cost. The update rule in Eq. (\ref{eq:watkins1989learningearning}) is replaced by
\begin{equation}\label{eq:n-step}
Q^{\operatorname{new}}(s_0, a_0) = (1-\alpha) Q(s_0, a_0) + \alpha \left(\sum_{i=0}^5 \gamma^i c_{a_i} + \gamma^6  \operatorname{arg\,min}_{a'} Q(s_7, a')  \right) .
\end{equation}
The subscripts on $s$, $a$ and $c$ denote the number of steps after $s_0$. To estimate the future expected discounted cost, the actual cost incurred is observed for six steps before bootstrapping with $Q(s_7, a)$. This reduces the bias that comes from an incorrect estimation of $Q$ for the future state. The value of $Q$ for state $s_0$ can only be estimated when $s_7$ is reached. We observe faster convergence with $n$-step Q-learning compared to regular $1$-step Q-learning Eq. (\ref{eq:watkins1989learningearning}).

The ideal solution to calculate the cost $c_a$ is to compute the weighted average of the cost in between the two bus stops $s$ and $s'$. This comes at a significant computational cost since, at every time step after taking action $a$, the cost is calculated and averaged until bus stop $s'$ is reached.
Other choices are possible, such as measuring $c_a$ just after taking the action $a$, but we noticed that this penalises disallowing boarding because the passengers left at the bus stop $s$ may have a strong contribution to the average waiting time. It effectively biases the buses towards acting as regular buses and the average waiting time is not optimal. Computing the cost $c_a$ after some time, such as when the bus reaches the midpoint between $s$ and next bus stop $s'$, eliminates this effect if in the meanwhile another bus boards passengers from $s$. This is our solution for this paper.

The learning and testing are carried out in a simulation where $T$ is divided into $312$ time steps per loop. The boarding/alighting time $l^{-1}$ is fixed at one time step. At a bus stop $i$, $k_i$ new passengers arrive every time step on average, either after $\floor{1/k_i}$ or $\ceil{1/k_i}$ time steps. Buses move between two consecutive bus stops in $T/M$ time steps. Fig. \ref{fig:kAkB} (b) compares the analytical results with those obtained by the simulation for regular and express buses. The analytical results implicitly consider a continuous number of passengers, creating potential discrepancies with the simulation. In our scenarios, the discrepancy in average waiting time between analytical predictions and simulation results are within $1.5\%$. The average waiting time for regular and express buses shown in Fig. \ref{fig:performance_comparison_rel} are the analytical values from Eqs. (\ref{eq:case1}) and (\ref{eq:case2}). Hard-coded regular and express buses in the simulation used for reinforcement learning lead to lower average waiting times by approximately $1\%$ due to the discretisation. To get more precise results for the plot in Fig. \ref{fig:kAkB} (b), $T$ is subdivided into 10000 time steps.

\section*{Acknowledgements}
This work was supported by the Joint WASP/NTU Programme (Project no. M4082189) and the DSAIR@NTU Grant (Project no. M4082418).

\bibliographystyle{unsrt}
\bibliography{main}

\end{document}